# Three-dimensional nucleation and growth of deformation twins in magnesium

Sangwon Lee[a], Michael Pilipchuk[b], Can Yildirim[c], Duncan Greeley[d], Qianying Shi[e], Tracy D. Berman[e], Adam Creuziger[f], Evan Rust[f], Carsten Detlefs[c], Veera Sundararaghavan[b], John E. Allison[e], Ashley Bucsek[a,e*]

[a]Department of Mechanical Engineering, University of Michigan, Ann Arbor, MI, USA
[b]Department of Aerospace Engineering, University of Michigan, Ann Arbor, MI, USA
[c]European Synchrotron Radiation Facility, Grenoble Cedex 9, FR
[d]Los Alamos National Laboratory, Los Alamos, NM, USA
[e]Department of Materials Science and Engineering, University of Michigan, Ann Arbor, MI, USA
[f]National Institute of Standards and Technology, Gaithersburg, MD, USA

*Corresponding author. Email: abucsek@umich.edu

**Abstract:**

At two-thirds the weight of aluminum, magnesium alloys have the potential to significantly reduce the fuel consumption of transportation vehicles. These advancements depend on our ability to optimize the desirable versus undesirable effects of deformation twins: three-dimensional (3D) microstructural domains that form under mechanical stresses. Previously only characterized using surface or thin-film measurements, here, we present the first 3D in-situ characterization of deformation twinning inside an embedded grain over mesoscopic fields of view using dark-field X-ray microscopy supported by crystal plasticity finite element analysis. The results reveal the important role of triple junctions on twin nucleation, that twin growth behavior is irregular and can occur in several directions simultaneously, and that twin-grain and twin-twin junctions are the sites of localized dislocation accumulation, a necessary precursor to crack initiation.

The use of lighter structural components can substantially reduce the fuel consumption of gas-powered vehicles and lower the energy density needs for battery-powered vehicles. These compelling material needs drive much of the interest in magnesium alloys. Magnesium is not only lightweight (weighing 30% less than aluminum) with high specific strength, but it is also the eighth most common element in the earth's crust (*1*). In comparison with other structural alloys such as Fe, Al, or Ti based alloys, there are opportunities to extract Mg from seawater, recycle Mg easily and/or biodegrade Mg alloys. These desirable properties make magnesium an environmentally attractive alternative to alloys currently used in the automotive sector.

The mechanical behavior of magnesium alloys, however, is complex. Magnesium and many of its alloys have a hexagonal close-packed (hcp) crystal structure. This structure causes strong plastic anisotropy and a limited number of dislocation slip systems, resulting in a propensity for deformation twinning (*2*). Deformation twins—three-dimensional (3D) microstructural domains that form under mechanical stresses—have a competing impacts on mechanical behavior. Deformation twins can accommodate significant plastic strain by enabling deformation along the c-axis of the crystal (*3*). However, deformation twins are often associated with crack initiation sites (*4–8*). Because of this dichotomy, twin nucleation, propagation, and growth mechanisms are critical for understanding the plasticity, ductility, and failure mechanisms of magnesium alloys.

While atomistic modeling and transmission electron microscopy (TEM) studies have shed light on the atomic-scale nature of deformation twinning (*7, 9–12*), recent studies emphasize the importance of 3D characterization for a more comprehensive understanding of deformation twinning at the grain scale (*11–13*). For example, grain-scale understanding of twin nucleation in the field relies largely on Schmid factors (*14*): a way of ranking twin variants in terms of likelihood based on the resolved shear stress on the twin plane in the twin shear direction. However, Lind et al used 3D high-energy diffraction microscopy (HEDM) to show that Schmid factors could not account for twin nucleation or variant selection behavior in pure zirconium, even when local grain orientations and stresses were included (*15*). Another HEDM study by Abdolvand et al showed that the most favored (highest Schmid factor) twin variant contributes the most to the twin number fraction, but the contribution of other variants is relatively independent of Schmid factor (*16*). In addition to twin nucleation, our understanding of twin propagation and growth has been stymied due to a lack of 3D *in-situ* characterization capabilities. Liu et al pointed out that the irregular shapes of deformation twins, especially their lateral and forward growth, cannot be fully understood without 3D analysis (*11, 12*). Their studies using statistical electron backscatter diffraction (EBSD) analysis and atomistic simulations also suggested that the lateral expansion of twins is faster than their forward propagation. This anisotropic growth behavior underlines the importance of 3D analysis for a complete understanding of twin evolution. Collectively, these findings emphasize the necessity 3D characterization techniques to better understand the complexities of twin behavior in magnesium alloys and twinning materials in general.

In the present work, we use dark-field X-ray microscopy (DFXM) to characterize in 3D deformation twinning inside an embedded grain in a bulk, polycrystalline Mg-4Al alloy with sub-micron spatial resolution. We directly observe the emergence and evolution of deformation twins *in situ* in connection to triple junctions and dislocation accumulation under applied stress conditions. The results reveal 3D twin growth behavior for the first time, confirming that twins in 3D appear as irregular ellipsoid shapes with its shortest axis in the twin plane normal $k_1$, its second longest axis in the twin shear direction $\eta_1$, and its longest axis in the twin lateral direction $\lambda$ (a vector perpendicular to the twin plane normal and the shear direction, $\lambda = k_1 \times \eta_1$). These results also show that growth along the lateral direction can persist to the later stages of twin growth, occurring either sequentially or simultaneously alongside growth in the twin shear direction and coarsening in the twin plane normal direction. The results also suggest a new preference for twin variant selection based on the geometry of triple junctions. Finally, we use the high angular resolution of DFXM combined with its high spatial resolution to reveal the local accumulation of geometrically necessary dislocations (GNDs)—a necessary precursor to crack initiation—at twin-grain junctions, twin-twin junctions, and along twin planes.

**Dark-Field X-Ray Microscopy Experiments**

DFXM is an X-ray diffraction microstructure imaging technique that can be used to image sub-surface grains with spatial resolutions as high as 60 nm (*17*). Because imaging is performed in the diffraction condition over large working distances, DFXM can be used to map distortions in elastic strain and orientation with extremely high angular resolutions, on the order of $10^{-5}$ and 0.001° (*18*), respectively. Together, these capabilities offer spatial resolutions approaching those of electron microscopy techniques with angular resolutions exceeding those of electron microscopy—but in 3D and over mesoscopic fields of view. Here, we use *in-situ* DFXM to measure the emergence and growth of extension twins and relate local orientation gradients to the accumulation of GNDs. The DFXM experiment was performed on beamline ID06-HXM at the European Synchrotron Radiation Facility.[1] The experimental setup is illustrated in **Fig. 1A**. The measurements use a 17 keV incident X-ray box beam that is approximately 250 μm wide and 200 nm tall (*18–21*). DFXM measurements were taken on a {001}-type Bragg reflection, which is magnified through a using a compound refractive lens (CRL) condenser objective consisting of 88 two-dimensional Be lenses. The sample-to-objective, objective-to-detector, and sample-to-detector distances were 279.6 mm, 4925.7 mm, and 5205.3 mm, respectively, resulting in a spatial resolution of 212 nm. The deformation twins are measured via diffraction contrast and identified from their twin plane normal. The relative misorientation of the parent grain is spatially resolved by tilting the sample about two vectors perpendicular to the scattering vector (sample tilting angles α and β) and calculating the centroids of the resulting intensity distributions at each pixel (*18–23*).

The mechanical loading was applied using a custom miniature loading stage developed at the National Institute of Standards and Technology (NIST) with a mass of 240 grams (**Fig. S4**). Sample geometry is shown in **Fig. S1**. A uniaxial tensile load is applied manually while the applied force was read from a load cell. A spring was used inline with the tensile load to absorb stress relaxation. The Mg-4Al alloy was provided by CanmetMATERIALS, Canada in the form of an extruded bar with a nominal mass fraction of 0.04 Al and balance Mg. The initial billet had a nominal diameter of 85 mm and length of 100 mm. The extrusion process reduced the nominal diameter to 15 mm, resulting in an extrusion ratio of 32.1. Process parameters for the billet and tooling were a temperature of 413°C during extrusion, extrusion speed of 254 mm/min. Annealing heat treatment at 515°C for 30 minutes was performed following the extrusion. The resultant textures from extrusion are shown in **Fig. S2**. Laboratory diffraction contrast tomography (labDCT) measurements were used to pre-characterize the 3D grain map prior to the DFXM experiment. These measurements, conducted using a Zeiss Xradia Versa 520 Micro-CT, were used to pre-select an embedded grain of interest with its c-axis aligned with the loading direction (**Fig. 1B**) to maximize twinning volume fraction (**Fig. S3**).[2]

---

[1] This microscope now resides at beamline ID03 at the European Synchrotron Radiation Facility (ESRF).

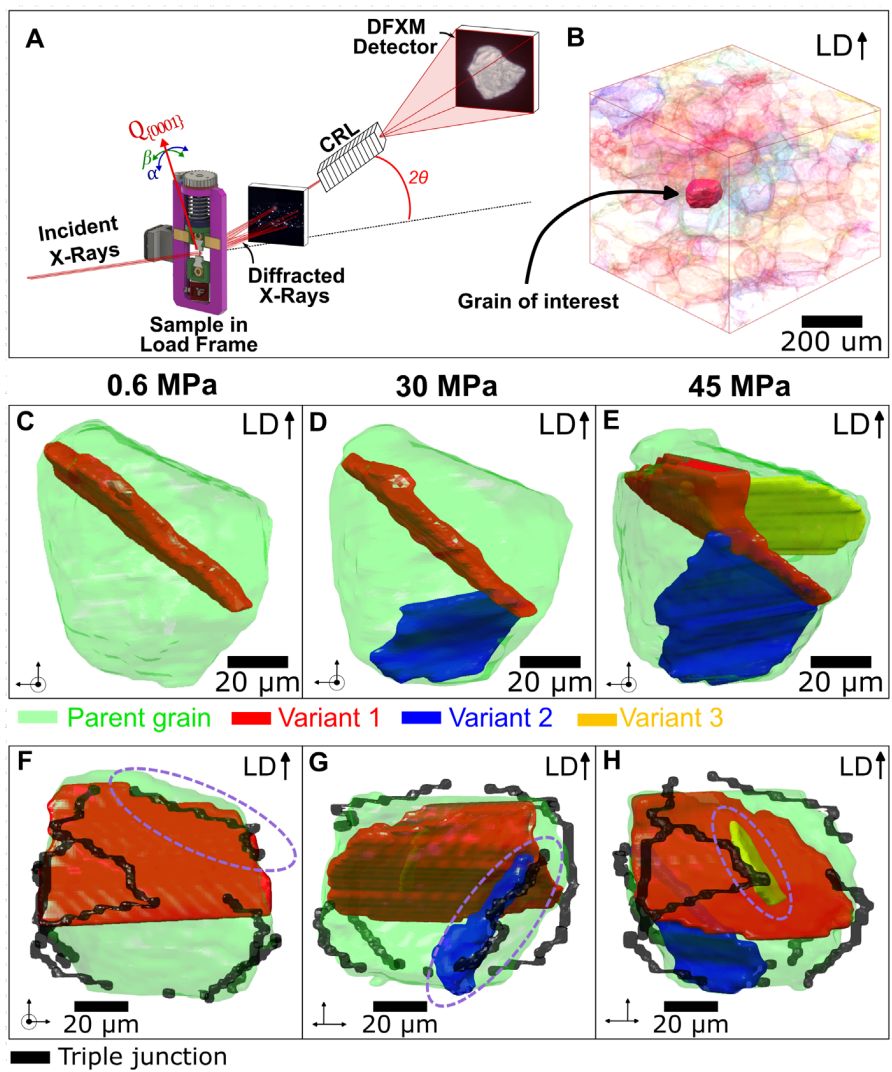

**Fig. 1.** *In-situ* **DFXM and 3D twin morphology measurements.** (**A**) *In-situ* DFXM technique; (**B**) Grain of interest inside the grain network; (**C−E**) Evolution of the twin morphology inside the parent grain; **F−H** shows the parent grain (green; 50% transparency), the triple junctions (opaque; black), and the three twins (opaque; red, blue, and yellow). The load direction (LD) is also marked. The purple dashed lines in **F−H** highlight the coincidence of twin-grain junctions to triple junction segments. See **Movies S1−S7** in the **Supplemental Material** for full 3D characterization.

---

[2] Certain equipment, instruments, software, or materials are identified in this paper in order to specify the experimental procedure adequately. Such identification is not intended to imply recommendation or endorsement of any product or service by NIST, nor is it intended to imply that the materials or equipment identified are necessarily the best available for the purpose.

## Results and Discussion

### *Twin Nucleation*

To study the twin nucleation behavior, we selected a grain that was oriented such that the basal plane was perpendicular to the loading direction (**Fig. 1B**). As a result, all six variants of the $\{10\bar{1}2\}$ extension twin system have twinning Schmid factors close to the maximal value of 0.5, with exact values ranging between 0.47 and 0.49. The neighboring grains had an average Schmid factor of 0.31 (**Fig. S6**). The high likelihood for this grain to twin relative to its neighbors makes this grain ideal for observing twin initiation events (rather than twin transfer across grain boundaries), while still subject to network features like grain boundaries and triple junctions.

**Fig. 1C−E** shows the 3D morphologies of the $\{10\bar{1}2\}$-type extension twins that formed during loading. The first twin that formed is a $(0\bar{1}12)[01\bar{1}1]$ twin (shown in red). This twin was not visible during the labDCT pre-characterization and we did not observe annealing twins in this material, so it likely formed when the sample was mounted and pre-loaded to 0.6 MPa. The second twin, a $(\bar{1}102)[1\bar{1}01]$ twin (shown in blue), was not observed until the 30 MPa load step. The third twin, a $(\bar{1}012)[10\bar{1}1]$ twin (shown in yellow), was not observed until the 45 MPa load step. Although all six $\{10\bar{1}2\}$-type extension twin variants have high Schmid factors, the three variants that formed had the lowest Schmid factors of the six variants (**Table S1**). This suggests that there may be additional factors contributing to twin variant selection beyond Schmid factor alone. **Fig. 1F−H** shows the twin morphologies relative to the locations of triple junctions (shown in black) measured using labDCT. The results (see purple circles) show that all three twin domains intersect triple junction segments. Furthermore, the results show that all three twin planes are geometrically aligned with triple junction segments, i.e., each twin plane lies parallel to the triple junction line segment that it intersects. This geometrical alignment between twin plane and triple junction is observed consistently, for all three twins.

While it is well known that triple junctions lead to stress concentrations and local stress concentrations can lead to twin nucleation events, the observation that the twin planes and triple junctions are geometrically aligned has not been made before to our knowledge. Twinned microstructures in bulk materials have been primarily characterized using 2D surface EBSD measurements, wherein such an observation would not be possible. These new observations suggest that the triple junctions are not only probable sites of twin nucleation, but the 3D geometry of triple junctions may also have an influence on twin variant selection, at least in the case where all twin variants have comparable Schmid factors. Specifically, these results suggest that twin variants with twin planes parallel to a triple junction line may be more favorable. To further explore this phenomenon, we conducted crystal plasticity finite element (CPFE) simulations using PRISMS-Plasticity (*24*) as detailed in the **Supplementary Material**, to understand stress evolution and its influence on twin nucleation during deformation. The simulations revealed that triple junction regions exhibit significantly higher stress concentrations compared to the surrounding microstructure, supporting their role as potential nucleation sites. To test this hypothesis, we measured the triple junction length that would be intersected by each of the six $\{10\bar{1}2\}$-type variants had they nucleated anywhere in the grain. The results (**Fig. S7**) show that the maximum possible triple junction intersection areas are the three variants that we observed, nucleated at the three locations we observed. This hypothesis may be supported by observations by Jiang et al that multiple twin nuclei can merge into a larger twin nucleus (*10*). In other words, multiple twin nuclei may form along a triple junction line and then merge, resulting in a twin domain that lies parallel to the triple junction line. This speculation is also in line with the observation that twin domains grow fastest/earliest in the twin lateral direction $\lambda$, which is perpendicular to the twin plane normal, discussed more in the next paragraph.

*Twin Growth*

**Fig. 2** shows the growth behavior observed for the three twins relative to the twin plane normal ($k_1$), twin shear direction ($\eta_1$), and twin lateral direction ($\lambda = k_1 \times \eta_1$). In (*11*), Liu et al used *ex-situ* EBSD statistical analysis on 2D twin shapes to suggest that twins grow faster along the lateral direction $\lambda$ than along the twin shear direction $\eta_1$ at the initial stages of twin growth, and that twins have irregular ellipsoid shapes suggesting that growth is locally dominated. In this study, we also observe that twins in 3D appear as irregular ellipsoid shapes, and we also observe that its longest axis is in the lateral direction $\lambda$, its second longest axis in the twin shear direction $\eta_1$, and the shortest axis in the twin plane normal direction $k_1$ (**Fig. 2A−F**). These observations are consistent for all three twins. Surprisingly, however, these results also show that growth along the lateral direction $\lambda$ can persist during the later stages of twin growth (**Fig. 2A**, **B**). In fact, these results show that the later stages of twin growth can involve growth in the lateral direction $\lambda$, growth in the twin shear direction $\eta_1$, and/or as coarsening in the twin plane normal $k_1$ direction, either sequentially or simultaneously. These observations are described in the next paragraph.

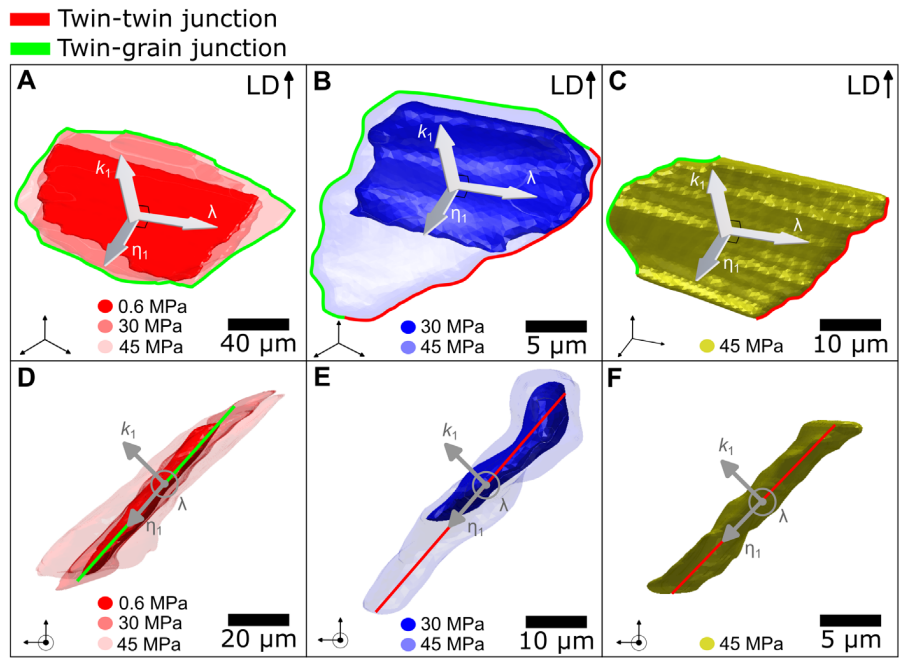

**Fig. 2. 3D twin propagation and coarsening.** Growth behavior of the three twins in (**A−C**) isometric viewing, and (**D−F**) with the viewing direction parallel to twin lateral vectors. The grain boundary is marked in **A−D** in green, and the red twin is marked in **B**, **C**, **E**, and **F** in red. See **Movies S8−S10** in the **Supplemental Material** for full 3D characterization.

The red twin (**Fig. 2A**, **D**) was the first twin to form. At 0.6 MPa, this twin has its longest axis in the lateral direction $\lambda$, but the domain had not completed growth in this direction yet (i.e., it had not yet reached the grain boundaries on either side, looking at **Fig. 2A**). Between 0.6 MPa and 30 MPa, instead of completing its growth in the lateral direction $\lambda$, the red twin grows in the twin shear direction $\eta_1$ (touching the bottom grain boundary in **Fig. 2A**), with very little growth in the lateral direction $\lambda$. (Note: The top edge of the red twin domain is the triple junction where we suspect initiation that was highlighted in **Fig. 1F**.) It is not until between 30 MPa and 45 MPa that the red twin completes its growth in the lateral direction $\lambda$ to reach the grain boundaries (on the left and ride edges, looking at **Fig. 2A**). During this same load step, the twin significantly and irregularly coarsened in the twin plane normal $k_1$ direction. The observations are similar for the blue twin (**Fig. 2B**, **E**), which was first observed at 30 MPa with its longest axis in the lateral direction $\lambda$. However, again, the twin does not grow far enough in this direction to touch the grain boundary (on the left edge, looking at **Fig. 2B**). Between 30 MPa and 45 MPa, the blue twin grows in all three directions simultaneously: It grows a small amount in the lateral direction $\lambda$ to reach the grain boundary, it grows a significant amount in the twin shear direction $\eta_1$ to reach the grain boundary on top and the grain/twin boundaries on bottom, and it coarsens irregularly in the twin plane normal $k_1$ direction.

One reason for the incomplete growth in the lateral direction $\lambda$ is the presence of negative backstresses. Backstress is the difference in twin resolved shear stress (TRSS) before and after the twin forms under the same applied strain. It has been suggested that backstress is important to understand, as it can either promote (if a positive backstress) or inhibit (if a negative backstress) subsequent twin formation and growth (*25*). In the **Supplemental Material**, we use CPFE simulations to investigate the backstress associated with the formation of the red twin (**Fig. 2A**, **D**). The results, shown in **Fig. S11**, indicate negative backstresses near the twin (that would inhibit growth) and positive backstresses near the grain boundaries (that would promote the formation and transmission of new twins). The experimental observations shown in **Fig. 2** combined with the insights provided by CPFE simulations underscore the importance of understanding backstresses and their possible role on twin formation and growth.

*Twinning and Dislocation Accumulation*

In addition to twin morphology, *in-situ* DFXM was used to spatially resolve local orientation gradients in the parent grain during deformation twinning. **Fig. 3** shows the evolution of intragranular misorientation (**Fig. 3D−F**), kernel average misorientation (KAM) (**Fig. 3G−I**), and GND density (**Fig. 3J−L**). Intragranular misorientation refers to the misorientation of each voxel with respect to the grain average. KAM refers to the misorientation of each kernel with respect to its neighbor kernel (kernel size: 5 μm). The voxels in these visualizations are opaque, so they only show the orientation gradients toward the grain exterior near the grain boundary. The dashed lines indicate where twin-grain junctions (i.e., where the twins intersect the grain boundary) are located. The results show that twin-grain junctions are sites of large local orientation gradients, these orientation gradients are connected to the accumulation of GNDs, and the GND density at these twin-grain junctions increases in severity with loading. These observations are described in the next paragraph.

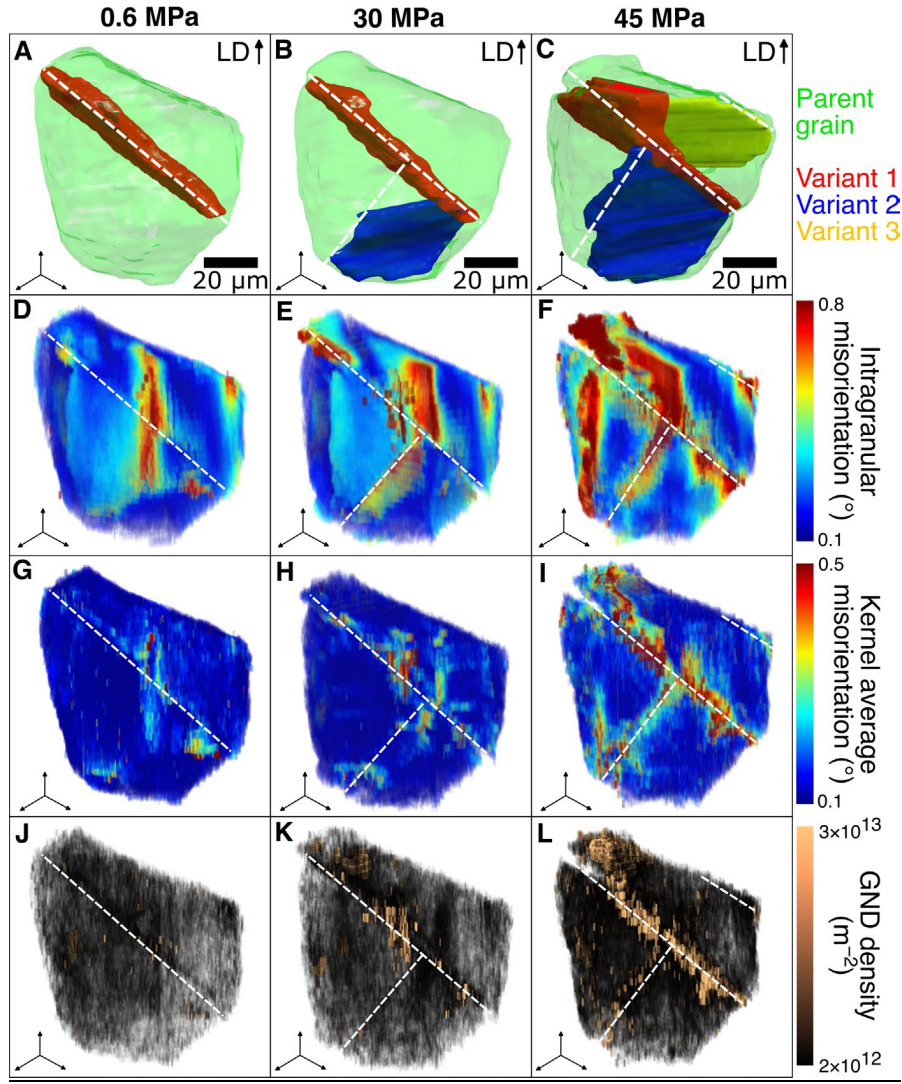

**Fig. 3. Evolution of local orientation gradients and GND density near the grain exterior.**
(**A−C**) Evolution of the 3D twin morphology, (**D−F**) intragranular misorientation with respect to the parent grain average, (**G−I**) KAM, (**J−L**) and GND density inside the parent grain. White dashed lines indicate the locations of twin-grain junctions. See **Movies S11−S19** in the **Supplemental Material** for full 3D characterization.

In the first measurement at 0.6 MPa (**Fig. 3G**), a faint line of high KAM can be seen along the red twin-grain junction. Between 0.6 MPa, 30 MPa (**Fig. 3H**), and 45 MPa (**Fig. 3I**), the KAM along this twin-grain junction increases from roughly 0.1° to 0.5°. The results are similar for the blue twin-grain junction, where a line of high KAM can be observed at 30 MPa (**Fig. 3H**). This line did not exist at 0.6 MPa when the blue twin-grain junction was not yet present, and it intensifies between 30 MPa (**Fig. 3G**) and 40 MPa (**Fig. 3I**). These local orientation gradient measurements were used to calculate GND density. One can see how regions of high KAM in **Fig. 3G−I** directly translate to the local accumulation GND density accumulation in **Fig. 3J−L**. The GND density is high at twin-grain junctions and low everywhere else, and the GND density increases at the twin-grain junctions as the macroscopic stress is increased. The observation that GNDs accumulate at twin-grain junctions is in line with theoretical work that suggest that the simple shear deformation associated with twinning cannot be applied to the region of the parent grain connected to a grain boundary (since it is fixed to the neighboring grain) (*26*).

Although twinning is necessary for hcp materials for strain compatibility and ductility, the negative effect of twinning is that twins lead to the development of crack initiation sites. Crack initiation sites develop when a region of high plastic deformation, or dislocation pileup, is adjacent to a region of high stress (*27*). While many computational tools have been used to theorize about the mechanisms behind twinning and crack initiation, e.g., (*28*), a fundamental understanding based on experimental evidence is lacking. The measurements presented in **Fig. 3** provide direct observations of how twins can lead to regions of high plastic deformation. Although we do not resolve local stresses in neighboring grains in this study, we know from numerous previous works that twinning can lead to stress concentrations across a grain boundary, particularly when there is not a geometrically aligned slip/twin system available in the neighboring grain (*29*). The results shown in **Fig. 3** suggest that twin-grain junctions will inevitably serve as sites for dislocation accumulation, at least in this material. When this dislocation accumulation is paired with a neighboring grain that is poorly aligned for slip/twin transfer, it can lead to adjacent stress concentrations across grain boundaries and, ultimately, crack initiation sites. Although there are numerous EBSD studies of twinned magnesium alloys, the orientation gradients and GND density accumulation at twin-grain junctions in bulk materials have, to our knowledge, not been experimentally observed before. We suspect that conventional EBSD does not sufficient angular resolution to resolve these features, and these new observations can be attributed to the superior orientation resolution (0.001°) of DFXM, combined with its high spatial resolution.

While it is clear from **Fig. 3** that GND density accumulates at twin-grain junctions, we also investigated if GND density accumulates at other features, like twin-twin junctions (i.e., where the twin domains intersect other twin domains), twin planes, and triple junctions. **Fig. 4** shows how local orientation gradients and GND density evolves inside the grain using a 2D "slice" of the grain (slice taken near the grain center). (All slices are available in **Movies S20−S28** in the **Supplemental Material**.) The intragranular misorientation results in **Fig. D−F** show that significant intragranular misorientation can be seen near grain boundaries and along twin planes. However, the local orientation gradients are much more concentrated in the KAM maps (**Fig. G−I**). There is only moderate GND density accumulation along twin planes, with slightly higher values at the twin-twin junctions, and no significant GND density was observed at triple junctions (except for the triple junction regions that also happened to be twin-grain junctions). In summary, the highest GND density accumulation was observed at twin-grain junctions with more moderate GND density accumulation at twin-twin grain junctions and along twin planes. These regions of local accumulation of GND density were significantly higher than anywhere else in the grain, including at triple junctions.

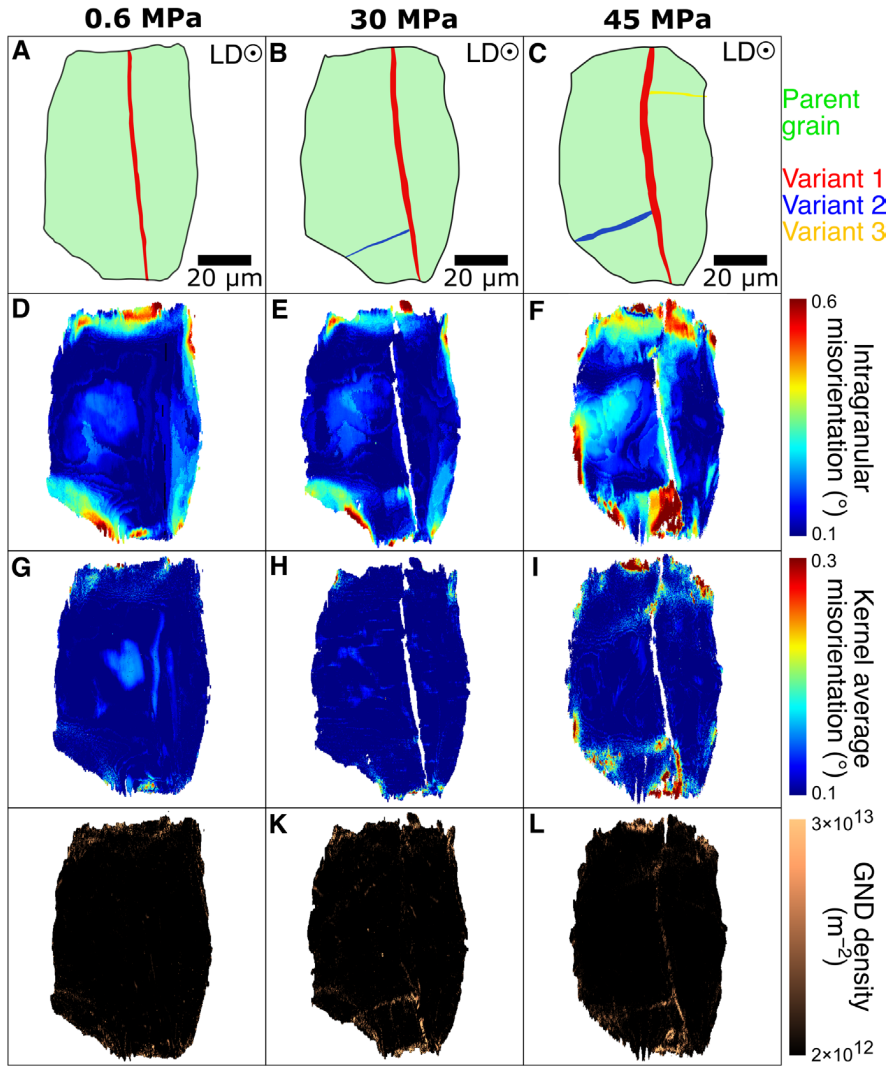

**Fig. 4. Evolution of local orientation gradients and GND density in the grain interior.**
(**A−C**) Evolution of the twin morphology, (**D−F**) intragranular misorientation with respect to the parent grain average, (**G−I**) kernel average misorientation, (**J−L**) and GND density inside the parent grain.

## Conclusion

In conclusion, we present the first 3D *in-situ* characterization of deformation twinning inside an embedded grain over mesoscopic fields of view using *in-situ* DFXM. The results confirm that twin growth behavior is irregular, leading to irregular 3D ellipsoidal shapes as previously suspected from EBSD measurements. While twin growth initially takes place primarily in the lateral direction, it can become stunted in this direction before traversing the grain. As a result, growth along the lateral direction can persist to the later stages of twin growth, occurring either sequentially or simultaneously alongside growth in the twin shear direction and coarsening in the twin plane normal direction. Using CPFE simulations, we show how the results may underscore the importance of understanding backstresses to predicting subsequent twin formation and growth. The results also reveal new observations regarding the 3D geometry of the twin planes with respect to the triple junctions, suggesting the geometry of triple junctions may influence twin variant selection. Finally, the results reveal that twin-grain junctions are the sites of localized GND accumulation, a necessary precursor to crack initiation. GNDs are also observed to accumulate to a lesser extent at twin-twin junctions, and to an even lesser extent along twin planes. The GND accumulation observed at these regions is substantially higher than anywhere else in the grain. These 3D *in-situ* observations offer new insights into deformation twinning and its relationship to nucleation, growth, and dislocations, contributing to a fundamental understanding of the complex micromechanical behavior of magnesium alloys and twinning materials in general. More broadly, this study highlights the importance of high-resolution 3D *in-situ* characterization to understanding 3D deformation mechanisms and microstructure evolution.

**Acknowledgments:**

**Funding:** U.S. Department of Energy Office of Basic Energy Sciences Division of Materials Science and Engineering Award #DE-SC0008637 as part of the Center for Predictive Integrated Structural Materials Science (PRISMS). (SL, MP, QS, DG, TB, VS, JA, AB)

**Author contributions:**

Conceptualization: SL, AB

Methodology: SL, AC, ER, CY, CD, AB

Visualization: SL

Funding acquisition: JA, AB

Supervision: AB

Material preparation and pre-characterization: QS, TB, SL

Simulations: MP, DG, VS

Experiments: SL, CY, CD, AB, AC, ER

Writing – original draft: SL, AB

Writing – review & editing: SL, CY, TB, AC, ER, CD, JA, AB

**Competing interests:** Authors declare that they have no competing interests.

**Data and materials availability:** All processed data are available to the public on the Materials Commons at https://doi.org/[*DOI to be established upon article acceptance*]. All raw data are available via the European Synchrotron Radiation Facility (ESRF) at https://doi.org/10.15151/ESRF-ES-901214768 (*30*).